\documentclass{article}
\usepackage[backref=page]{hyperref}
\usepackage{graphicx, float, amsmath,cleveref, array, caption}
\usepackage{wrapfig}
\usepackage{multirow}
\usepackage[numbers,compress,sort]{natbib}
\usepackage{parskip}
\usepackage{xcolor}
\usepackage{booktabs}
\usepackage{appendix}
\usepackage{authblk}
\title{Data-Driven Bifurcation Handling in Physics-Based Reduced-Order Vascular Hemodynamic Models}
\author[1]{Natalia L. Rubio}
\author[1, 2]{Eric F. Darve}
\author[1, 2, 3,4* ]{Alison L. Marsden}
\affil[1]{Stanford University - Department of Mechanical Engineering}
\affil[2]{Stanford University - Institute for Computational and Mathematical Engineering}
\affil[3]{Stanford University - Department of Bioengineering}
\affil[4]{Stanford University - Department of Pediatrics - Cardiology}
\affil[*]{Corresponding author: amarsden@stanford.edu}
\date{}                     
\usepackage[total={6in,9in}]{geometry}
\Crefname{equation}{Eq.}{Eqs.}

\begin{document}
\maketitle
\textbf{Background and Objective:}

Three-dimensional (3D) computational fluid dynamics simulations of cardiovascular flows provide high-fidelity hemodynamic predictions to support cardiovascular medicine,  but require substantial computational resources, limiting their clinical applicability. Reduced-order models (ROMs) offer computationally efficient alternatives but suffer from significant accuracy losses, particularly at vessel bifurcations where complex flow physics are inadequately captured by standard Poiseuille flow assumptions. This work presents an enhanced numerical framework that integrates machine learning-predicted bifurcation coefficients into 0D hemodynamic solvers to improve accuracy while maintaining computational efficiency. 

\textbf{Methods:}

We develop a resistor-resistor-inductor (RRI) model that uses neural networks to predict pressure-flow relationships from bifurcation geometry, incorporating both linear and quadratic resistance terms along with inductive effects. The method employs physics-based non-dimensionalization to reduce training data requirements and includes flow split prediction for improved geometric characterization. We incorporate the RRI model into a zero-dimensional (0D) cardiovascular flow model using an optimization-based solution strategy. We validate the approach in isolated bifurcations and vascular trees containing up to 40 junctions across Reynolds numbers ranging from 0 to 5,500, defining ROM accuracy by comparison to high-fidelity 3D finite element simulation results. 

\textbf{Results:}

Results demonstrate substantial accuracy improvements: averaged across all trees and all Reynolds numbers, the RRI method reduces inlet pressure errors from 54 mmHg (45\%) for standard 0D models to 25 mmHg (17\%), while a simplified resistor-inductor (RI) variant achieves 31 mmHg (26\%) error. The enhanced 0D models show particular effectiveness at high Reynolds numbers and in extensive vascular networks. 

\textbf{Conclusions:}

This hybrid numerical approach enables accurate, real-time hemodynamic modeling suitable for clinical decision support, uncertainty quantification, and digital twin applications in cardiovascular biomedical engineering.

\section{Background and Objective}

In recent decades, numerical models of cardiovascular flows have become increasingly influential tools in clinical management of cardiovascular disease (including diagnosis, risk stratification, and surgical planning) \cite{Bluestein2017UtilizingClinic, Morris201ComputationalMedicine, Schwarz2023BeyondDisease}, medical device design \cite{Farah2025ComputationalAneurysms, Hsu2025ConstrainedGrafts, Blum2025OversizedGrafts, Hu2025MultiphysicsConduit, Fraser2011TheDevices, Frank2002ComputationalDesign, Gundert2012OptimizationDynamics, hossain12mathematical, Kung2013PredictiveCases}, and the study of fundamental biological mechanisms driving disease progression \cite{Brown2024ComputationalDiseases, Rolf-Pissarczyk2025MechanismsModels, Baumler2024LongitudinalDynamics,Baumler2025AssessmentModels, Dong2021ComputationalDefects}.  Traditionally, pressure and velocity in the vasculature are determined by solving the Navier-Stokes equations using high-fidelity, 3D finite-element techniques that require significant computational effort.  Reduced-order models (ROMs) present a computationally lightweight alternative, solving simplified versions of the vascular flow in a fraction of the time. A zero-dimensional (0D) ROM can resolve a vascular flow in fractions of a second on a personal laptop, while 3D simulations generally take hours and require high-performance computing resources. As such, ROMs are valuable tools in cardiovascular modeling in cases where the computational load would be intractable using 3D simulation alone, namely, for extensive vasculatures, in many-query applications such as optimization and uncertainty quantification (UQ), and for real-time computation as needed in digital twins.

First, ROMs are commonly used to represent vasculatures too extensive to model in 3D.  Often, they are used as boundary conditions (BCs), where they represent vasculatures downstream of the anatomy of interest whose behavior must be taken into account but would be too costly to simulate in 3D.\cite{Olufsen1999StructuredArteries, Kheyfets2013ConsiderationsHypertension, Hunter2010ComputationalHypertension, Johnson2011ApplicationPredictions, Clipp2009ImpedanceRespiration, Kheyfets2015Patient-specificCirculation, Ebrahimi2022SimulatingModel, Cai2024ACapacity, Brown2023AMechanics}.  Often, these distal vessels are too small to be captured in medical images - in these cases, realistic synthetic vasculatures may be generated and modeled with ROMs to produce more physiological BCs \cite{Sexton2025RapidBiomanufacturing, Menon2024PersonalizedTrees, Talou2021AdaptiveProcesses, Kerautret2023OpenCCO:Trees}.  In some cases, ROMs are also used in closed-loop circulatory physiology models, sometimes including cardiac and organ compartments \cite{Seresti2025ValidationMeasurements, Hu2025MultiphysicsConduit, Menon2024PersonalizedQuantification, Kim2010Patient-specificArteries, Augustin2021ACirculation,  Cousins2012BoundaryRevisited, Mynard2008AMethod, Brown2023RecentModeling}.  ROMs are not only used as boundary conditions, but also as stand-alone models, generally for large vascular systems that would be expensive to model in 3D \cite{Muller2014ASystem, Zhang2016DevelopmentData, Stergiopulos1992ComputerA, Charlton2019ModelingIndexes, Steele2003InGrafts, Sherwin2003ComputationalSystem, Pfaller2024ReducedModeling, Fullana2009ANetworks, Huberts2012AClinic, Qohar2021ASystem, Reymond2009ValidationTree, Mynard2008AMethod}. As biomanufacturing technology advances, the aforementioned synthetic vasculatures can be used to perfuse the printed material, enabling the fabrication of larger, more advanced tissues and organs.  ROMs also support the design of synthetic vasculatures that achieve desired flow and pressure targets by providing estimates of flow and pressure in the network \cite{Sexton2025RapidBiomanufacturing, Talou2021AdaptiveProcesses, Kerautret2023OpenCCO:Trees}

Second, ROMs are instrumental for many-query applications that require a large number of simulations that would be intractable in 3D.  One example is BC tuning for 3D simulations, which requires many model evaluations to identify BC parameters for which simulation results match clinical targets.  In the tuning process, ROMs can act as surrogates for the the 3D model to reduce the computational cost of evaluating many candidate parameter sets \cite{Richter2025BayesianModels, Brown2023AMechanics, Nair2023Non-invasiveMeasurements, Nair2023HemodynamicsSimulation, Li2023AFlow, Ramazanli2025ModelingApproaches}.  In recent work, ROM solutions were used to initialize 3D simulations, dramatically reducing the computation needed to arrive at a converged 3D solution \cite{Nair2023Non-invasiveMeasurements, Pfaller2021OnSimulations}.  Finally, because they are so lightweight, ROMs enable multifidelity uncertainty quantification techniques to achieve variance reduction at a fraction of the computational cost \cite{Fleeter2020MultilevelHemodynamics, Choi2025OnConditions, Tran2017AutomatedSimulations, Seo2020Multi-FidelityUncertainty, Seo2020TheWalls, Menon2024PersonalizedQuantification, Schiavazzi2018MULTIFIDELITYHEMODYNAMICS, Schaefer2024GlobalHaemodynamics, Zanoni2024ImprovedTechniques}.

Third, ROMs are increasingly enabling real-time cardiovascular flow modeling towards digital twins.  For instance, they can provide real-time estimates of flow, pressure, and clinically relevant markers, such as fractional flow reserve from patient-specific vascular geometries, and even to predict the effects of hypothetical modifications to the vasculature \cite{Pham2022SvMorph:Anatomies, Sankaran2020PhysicsSimulations, Hu2023AArteries, Hashemi2022RealReserve}.  They also support control systems for experimental studies of cardiovascular flows \cite{Nair2025ExperimentsCoarctations, Nestler2014AHeart, Petrou2018StandardizedStudy, Ochsner2013ADevices}.

ROMs can generally be divided into two classes: physics-based ROMs, which solve simplified, but analogous physical systems, and data-driven ROMs, which use machine learning techniques to capture and recreate hemodynamic behavior from data.  The primary drawback of classical physics-based ROMs is their limited accuracy, a consequence of the simplifications and assumptions made in deriving a lower-dimensional representation of the 3D flow.  In recent years, as machine learning methods have advanced and cardiovascular flow simulation data has become more accessible, data-driven ROMs have emerged as promising alternatives to their physics-based counterparts, achieving high accuracy at significantly lower computational cost than 3D simulations \cite{Arzani2021Data-drivenOpportunities, Pegolotti2024LearningNetworks, Pegolotti2021ModelVessels, Tenderini2025ModelInteraction, Ye2024Data-drivenSnapshots, Balzotti2022AGraft, Fresca2021Real-TimeModels, Liang2020AAorta, Gharleghi2022TransientNetworks, Li2021PredictionLearning}. Data-driven ROMs, however, may suffer from a lack of generalizability or interpretability, which are key for clinical adoption.

In this work, we investigate a data-informed bifurcation-handling method for use in physics-based ROM frameworks, aiming to capitalize on the accuracy gains made by data-driven techniques while retaining the simplicity and interpretability of physics-based ROMs.  A particularly challenging aspect of physics-based ROM development is the accurate numerical representation of vessel bifurcations, where standard Poiseuille flow assumptions break down due to complex three-dimensional flow phenomena, including flow separation, secondary flows, and pressure recovery. These effects cannot be captured by traditional lumped-parameter approaches, leading to significant numerical errors that propagate throughout the vascular network solution. The challenge is compounded in iterative solvers where bifurcation nonlinearities can cause convergence difficulties and increased computational overhead.

In this work, we present a hybrid numerical framework that combines machine learning with physics-based reduced-order modeling to address bifurcation handling challenges in cardiovascular flow simulation. We develop an enhanced resistor-resistor-inductor (RRI) numerical model that uses neural networks to predict pressure-flow coefficients from geometric features, extending previous work \cite{Rubio2025HybridDifferences} with novel algorithmic contributions including: (1) physics-based non-dimensionalization to improve numerical scaling and reduce training data requirements, (2) integrated flow split prediction for enhanced geometric characterization, and (3) a simplified resistor-inductor (RI) variant for computational efficiency. Our numerical validation employs high-fidelity 3D finite element simulations across Reynolds numbers from 0 to 5500, testing isolated bifurcations and vascular networks containing up to 40 junctions. We introduce an optimization-based solution strategy to handle the increased nonlinearity in the system of equations governing the 0D model using the RRI model. The resulting hybrid approach maintains the interpretability and computational speed of physics-based ROMs while achieving substantial accuracy improvements through data-informed numerical corrections, enabling real-time hemodynamic modeling suitable for clinical applications and uncertainty quantification frameworks.

\section{Models and Methods}
\subsection{ROM fundamentals}
The most common physics-based hemodynamic ROMs are one-dimensional (1D) models, which resolve flow and pressure along the vessel centerlines or 0D models, which resolve pressure only at the vessel inlets and outlets. 
Here, we focus on 0D ROMs, or lumped-parameter models, which represent a vasculature as an electric circuit, where pressure $P$ is analogous to voltage and flow $Q$ is analogous to current \cite{Formaggia2010CardiovascularSystem, Pfaller2022AutomatedFlow, Mirramezani2019ReducedArteries, Kim2010DevelopingFlow, Menon2025SvZeroDSolver:Simulations}.  Each vessel is represented by circuit elements whose characteristic values are determined from physics-based heuristics.  Typically, these are resistors $R$, inductors $L$, and capacitors $C$, representing viscous pressure loss in fully developed pipe flow, deformation of the vessel walls, and inertial effects, respectively.  Sometimes, a nonlinear resistor $R_\text{stenosis}$ is also used to capture flow separation effects in stenoses or aneurysms.  The equations relating the flow (current) and pressure (voltage) over each vessel are then
 \begin{equation}
     Q_\text{in}-Q_\text{out} = C (\dot{P}_\text{in} + R\dot{Q}_\text{in} + 2R_\text{stenosis}|Q_\text{in}|\dot{Q}_\text{in}),
     \label{eq:standard_mass_conservation}
 \end{equation}
and
\begin{equation}
    P_\text{in} - P_\text{out} = R Q_\text{in} + R_\text{stenosis} Q_\text{in} |Q_\text{in} | + L\dot{Q}_\text{out}.
    \label{eq:standard_mom_conservation}
\end{equation}
Vessel resistances and inductances are estimated from a fully-developed Poiseuille flow assumption, 
\begin{equation}
    R = \frac{8 \pi \mu l}{A^2},\;  L = \frac{\rho l}{A}, 
    \label{eq:physics_based_char_vals}
\end{equation}
where $A$ and $l$ are the cross-sectional area and length of the vessel, and $\mu$ and $\rho$ are the viscosity and density of blood.  For the rigid wall simulations considered in this work, the capacitance $C$ is assigned a very small value, $\mathcal{O}(10^{-8})\;\text{cm}^3 \; \text{Ba}^{-1}$, included only for numerical stability.  The stenosis resistance $R_\text{stenosis}$ is given by 
\begin{equation}
    R_\text{stenosis} = \frac{K_t \rho}{2A^2} \left( \frac{A}{A_\text{stenosis}} - 1\right)^2,
\end{equation}
where $K_t$ is an empirically fitted coefficient \cite{Steele2003InGrafts, Mirramezani2019ReducedArteries, Itu2013Non-invasiveMeasurements}.  

In the standard 0D model, we enforce conservation of flow and assume zero pressure drop in vessel bifurcations as
 \begin{equation}
     Q_\text{in}- \sum_\text{outlets}Q_\text{out} = 0,
 \end{equation}
and
\begin{equation}
    P_\text{in} - P_\text{out} = \Delta P = 0, \; \forall \text{ outlets}.
\end{equation}
1D models are another important family of ROMs that resolve the hemodynamics along the vessel centerlines, offering one degree of spatial resolution and higher fidelity compared to 0D models.  1D ROMs project the 3D Navier-Stokes equations onto the vessel centerlines by assuming a variable-area vessel cross-section and a known velocity profile.  A constitutive law describes the relationship between pressure and vessel deformation, which, along with the 1D conservation of mass and momentum equations, governs the pressure, flow, and cross-sectional area along the centerlines \cite{Ghigo2018AModel, Canic2003MathematicalVessels, Hughes1973OnVessels, Olufsen20045.Arteries, Wan2002ADisease, Wang2015VerificationModel, Wang2016FluidModel}.  

Like 0D models, 1D models often assume constant pressure over bifurcations \cite{Taylor-LaPole2023APatients, Olufsen1999StructuredArteries, Stergiopulos1992ComputerA, Reymond2009ValidationTree}, however some have accounted for pressure recovery effects  \cite{Fullana2009ANetworks, Sherwin2003ComputationalSystem, Lee2016MultiphysicsCHeart, Alastruey2009ModellingSynthesis, Alastruey2012ArterialHaemodynamics, Mynard2008AMethod}.  We focus on the 0D model in this work because it does not account for pressure recovery or other bifurcation pressure effects, and therefore stands to gain the most from the incorporation of a bifurcation handling model.  We note, however, that our developments may also prove beneficial in 1D models in future studies. While some studies have considered alternative physics-based and empirical models for pressure drop with mixed results, the standard handling described above remains the most common approach \cite{Mynard2015AJunctions, Chnafa2017ImprovedDrops, Mirramezani2020AFlow, Pewowaruk2021AcceleratedSwine, Blanco2018ComparisonReserve}.

\subsection{RRI method}
We recently introduced the RRI model \cite{Rubio2025HybridDifferences}, which models the pressure drop between the junction inlet and outlet as 
\begin{equation}
    \Delta P_\text{RRI} = R_{\text{lin}}(\mathcal{G}) Q + R_{\text{quad}}(\mathcal{G}) Q^2 + L(\mathcal{G}) \dot{Q}.
    \label{eq:RRI_model}
\end{equation}
The coefficients $R_{\text{lin}}$, $R_{\text{quad}}$, and $L$ are predicted from a vector $\mathcal{G}$ describing bifurcation geometry using a neural network.  Similarly to the circuit elements used to describe vessels in 0D models, $R_\text{lin}$ describes pressure losses proportional to flow, and $L$ represents inertial pressure losses proportional to the time-derivative of the flow.  The quadratic resistance $R_\text{quad}$ takes a slightly different form than the vessel stenosis resistance, proportional to $Q^2$ rather than $Q|Q|$.  The quadratic resistor aims to capture pressure recovery effects that are independent of the flow direction, while the stenosis resistor captures pressure losses associated with flow contraction and expansion, dependent on the flow direction.  In this study, we only consider rigid-wall vasculatures and therefore do not include a capacitance.

One of the improvements to the RRI model we include in this work is the inclusion of the flow split, $\phi$ in the geometry vector $\mathcal{G}$ along with inlet and outlet areas $A$, outlet lengths $l$, and outlet angles $\theta$, shown in \Cref{fig:junction_schematic}.  With this modification, 
$$\mathcal{G}_1 = \left[ A_\text{inlet},A_\text{1}, A_\text{2},l_\text{1},l_\text{2}, \theta_\text{1},\theta_\text{2}, \phi_1 \right],$$ 
where the subscripts 1 and 2 refer to the different outlets.

For comparison, we also consider a simplified resistor-inductor (RI) model, 
\begin{equation}
    \Delta P_\text{RI} = R_{\text{lin}}(\mathcal{G}) Q + L(\mathcal{G}) \dot{Q}.
    \label{eq:RI_model}
\end{equation}

\subsection{Finding RRI Coefficients}
\subsubsection{Physics-Based Scaling}
Expanding on \cite{Rubio2025HybridDifferences}, we introduce a physics-based scaling that yields a non-dimensional version of the bifurcation geometry to which machine learning techniques are later applied.  The physics-based scaling enables us to represent identically shaped bifurcations of different sizes with a single set of non-dimensional parameters, dramatically reducing the amount of training data needed to represent the vast range of bifurcation sizes encountered in vascular trees.

We choose the inlet radius as a characteristic length, $l_c$, which defines the scale of the geometry.  Given $l_c$ we define a characteristic area, $A_c$, velocity, $U_c$, time, $t_c$, and pressure, $P_c$:
\begin{equation}
    A_c = \pi l_c^2, \qquad
    U_c = \frac{\text{Re}_c \, \mu}{\rho (2l_c)}, \qquad
    Q_c = A_c U_c, \qquad
    t_c = \frac{l_c}{U_c}, \qquad
    P_c = \rho U_c^2.
    \label{eq:characteristic_values}
\end{equation}
We choose $\text{Re}_c$ = 4,500, a typical aortic Reynolds number at systole \cite{Ku1997BloodArteries, Stalder2011AssessmentMRI}, but highlight that the choice of $\text{Re}_c$ is arbitrary as the RRI coefficients are re-dimensionalized using the same $\text{Re}_c$. Given these characteristic values, we define non-dimensional variables $A^*$, $U^*$, $Q^*$, $t^*$, and $P^*$ as follows:
\begin{equation}
    A = A_c A^*, \qquad 
    U = U_c U^*, \qquad 
    Q = Q_c Q^*, \qquad 
    t = t_c t^*, \qquad 
    P = P_c P^*
    \label{eq:dimensionless_values}
\end{equation}
Substituting \Cref{eq:dimensionless_values} into \Cref{eq:pressure_continuity} gives
\begin{equation}
\Delta P^* P_c = R_{\text{lin}} Q_c Q^* + R_{\text{quad}} Q_c^2 Q^{*2} + L \frac{t_c}{Q_c}\dot{Q^*}.
    \label{eq:sub_nd_params_in}
\end{equation}
Rearranging yields
\begin{equation}
\Delta P^* = \underbrace{\frac{R_{\text{lin}} Q_c}{P_c}}_{R_{\text{lin}}^*} Q^* + \underbrace{\frac{R_{\text{quad}} Q_c^2}{P_c}}_{R_{\text{quad}}^*} Q^{*2} + \underbrace{\frac{L Q_c}{t_c P_c}}_{L^*} \dot{Q^*},
    \label{eq:rri_nd}
\end{equation}
where we define the coefficients multiplying $Q^*$, $Q^{*2}$, and $\dot{Q^*}$ as non-dimensional linear resistance, $R_{\text{lin}}^*$, quadratic resistance, $R_{\text{quad}}^*$, and inductance, $L^*$, respectively.

\begin{figure}[htbp]
    \centering
    \includegraphics[width=0.9\linewidth]{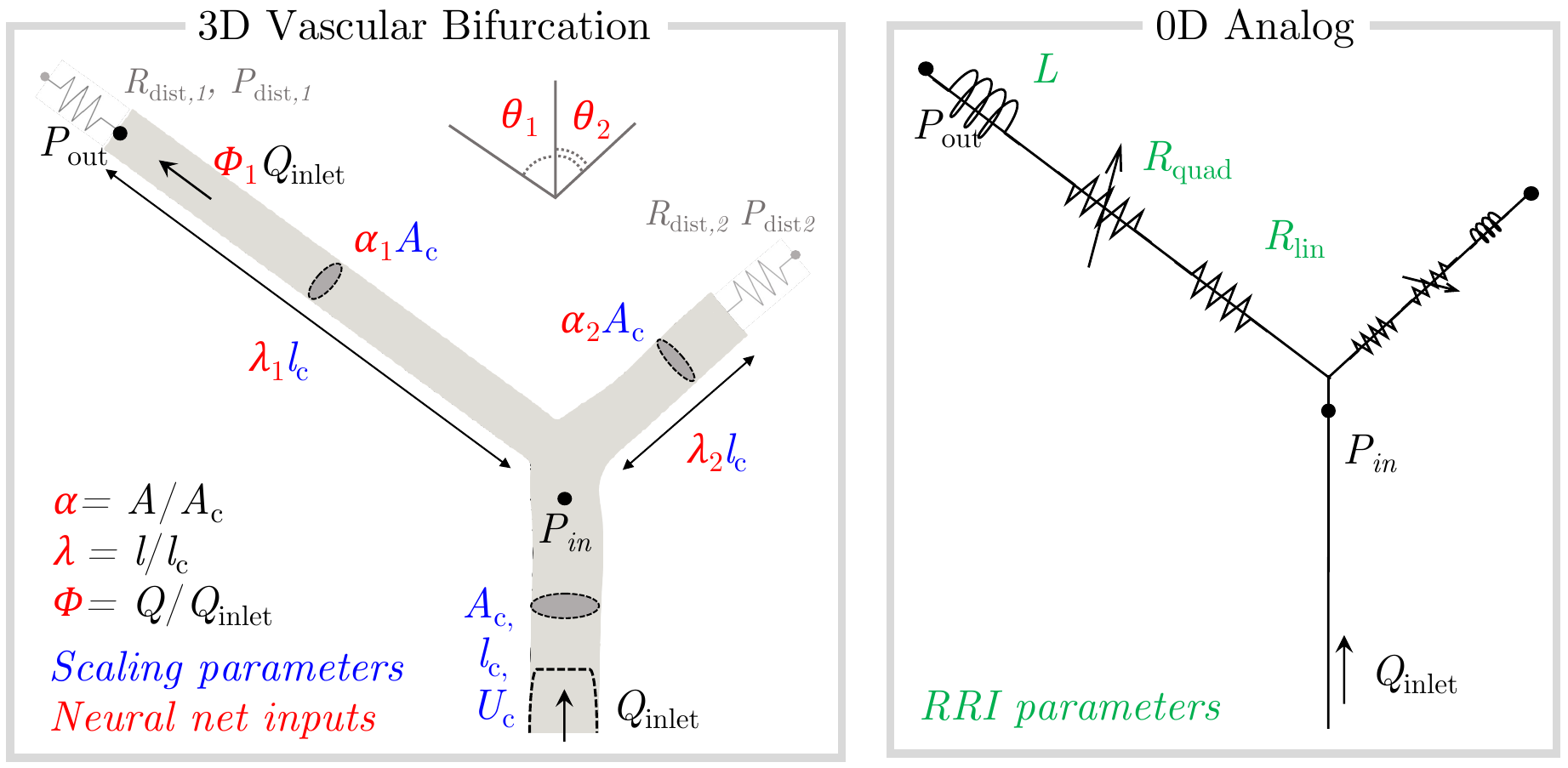}
    \caption{3D bifurcation, labeled with relevant scaling factors and the non-dimensional parameters used to predict the RRI parameters (right).  0D RRI representation of bifurcation using circuit components with characteristic parameters predicted from the 3D geometry.}
    \label{fig:junction_schematic}
\end{figure}

We predict $R_{\text{lin}}^*$, $R_{\text{quad}}^*$, and $L^*$ from a vector $\mathcal{G}^* = \left[\alpha_1, \alpha_2, \alpha_1^{-2}, \alpha_2^{-2},l_1, l_1^2,\theta_1, \theta_2, \phi_1, \phi_1^{-1} \right]$ containing non-dimensional geometric parameters shown in \Cref{fig:junction_schematic}.  They are easily computed from the entries of $\mathcal{G}$ as shown in \Cref{tab:non_dim_params}.

\begin{table}[h]
\renewcommand{\arraystretch}{1.5} 
\begin{tabular}{>{\raggedleft\arraybackslash}p{5cm} >{\raggedright\arraybackslash}p{3cm} >{\raggedright\arraybackslash}p{3cm}}
  normalized outlet areas:  & $\displaystyle \alpha_1 = A_{\text{outlet 1}}/{A_c}$, & $\displaystyle \alpha_2 = A_{\text{outlet 2}}/{A_c}$, \\
  normalized outlet lengths: & $\displaystyle \lambda_1 = l_{\text{outlet 1}}/{l_c}$, & $\displaystyle \lambda_2 = l_{\text{outlet 2}}/{l_c}$, \\
  outlet angles: & $\displaystyle \theta_1$, & $\displaystyle \theta_2$, \\
  flow ratio: & \multicolumn{2}{l} {$\displaystyle \phi_1 $  }\\
\end{tabular}
\caption{Non-dimensional geometric parameters from which RRI and RI coefficients are determined.  We estimate the flow split $\phi$ a priori from the downstream anatomy using \Cref{eq:flow_split} and  \Cref{eq:resistance_recursion}.
}
\label{tab:non_dim_params}
\end{table}

\begin{figure}[htbp]

    \begin{centering}
            \includegraphics[width=0.95\linewidth]{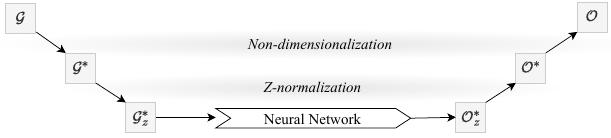}
    \caption{Data flow schematic for prediction of RRI coefficients from bifurcation geometry.}
    \label{fig:geo_to_rri}
    \end{centering}
\end{figure}

A Z-normalization is applied to the dimensionless geometry vector ${\mathcal{G}^*}$ such that the distribution of each element has a mean of zero and a standard deviation of one  
\begin{equation}
    \mathcal{G}_{z,i}^* = \frac{(\mathcal{G}^*_i - \mu_i)}{\sigma_i},
\end{equation}
where the parameters $\mu_i$ and $\sigma_i$ for each parameter are determined from the training set. This facilitates the optimization needed to train the neural network.

\subsubsection{Neural Network Model}
We use neural networks to predict the Z-normalized, dimensionless RRI and RI coefficients, $\mathcal{O}_z^*$, from the Z-normalized, dimensionless geometry vector, $\mathcal{G}_z^*$.  For each coefficient, $c$, we use a separate neural network of the form
\begin{align*}
    y_0 & = \text{ReLU}(W_{0, c} \mathcal{G}^*_z + b_{0, c}) \\
    y_i & = \text{ReLU}(W_{i, c} y_{i-1} + b_{i, c}) \;\;\; \forall i = 1...n-1 \\
   \mathcal{O}_{z, c}^* & = W_{n, c} y_{n-1} + b_{n, c} ,
\end{align*}
where $W$ and $b$ are weights and biases learned by the neural network.  The number of hidden layers, $n-1$, and the layer width are detailed in \Cref{app:nn_params}.  $W$ and $b$ are optimized by minimizing an objective function, 

\begin{equation}
    \text{Loss} = \left \lVert \mathcal{O}^*_{z,j, pred} - \mathcal{O}^*_{z,j, true} \right \rVert_2,
\end{equation}
over randomized batches of data with an Adam optimizer.  We use a dataset of approximately 800 bifurcations, with 90\% allocated for training and 10\% allocated for validation, and a batch size of 50.

As shown in \Cref{fig:geo_to_rri}, the output of the neural network, $\mathcal{O}^*$, is mapped back to physical RRI coefficients by inverting the z-normalization, 
\begin{equation}
    \mathcal{O}_{i}^* = \mathcal{O}_{i,z}^* \sigma_i + \mu_i
\end{equation}
and recovering the dimensionality of the RRI coefficients according to \Cref{eq:rri_nd},
\begin{equation*}
    \mathcal{O}^* = \begin{bmatrix}
    R_\text{lin}^* = R_\text{lin}\frac{Q_c}{P_c}, & R_\text{quad}^*=R_\text{quad}\frac{Q_c^2}{P_c},&  L^* = L \frac{t_c}{Q_c}
    \end{bmatrix}.
\end{equation*}

\subsubsection{Training Data Generation}
\label{sec:data_generation}
To train and validate the neural network, we generate a synthetic cohort of bifurcations representative of those encountered in vascular trees.  Each bifurcation is constructed based on its dimensionless geometry vector, $\mathcal{G}^*$, whose entries are chosen via Latin hypercube sampling from the parameter ranges observed in synthetic vascular trees and listed in \Cref{app:feature_ranges} \cite{McKay1979ACode}.

Assuming an inlet area of 1 cm$^2$, the rest of the geometry is defined by the area ratios $\alpha$ and angles $\theta$ prescribed in $\mathcal{G}$.  Using the SimVascular Python API \footnote{\url{https://simvascular.github.io/documentation/python_interface.html} (June 2025)}, the 3D geometry is constructed and meshed \cite{Updegrove2017SimVascular:Simulation}. \textit{(Note: this procedure is further detailed in \cite{Rubio2025HybridDifferences}).}  Next, we simulate flow through the bifurcation using \texttt{svMultiPhysics}  \footnote{\url{https://github.com/SimVascular/svMultiPhysics}}, an open-source 3D finite-element solver designed to simulate cardiovascular mechanics.  At the bifurcation inlet, we impose a time-varying inlet flow, $Q_\text{inlet}(t)$,  representing the systolic portion of the cardiac cycle, when pressure differences a generally most significant.  The inlet flow over time can be seen in \Cref{fig:isolated_fit}.  The maximum Reynolds number applied at the inlet is $\approx$5,500, chosen to match the maximum Reynolds number observed in the aorta \cite{Stalder2011AssessmentMRI, Cheng2025CharacteristicsSimulation, Mahalingam2016NumericalArteries, Fischer2007SimulationFlows}.  We prescribe a plug velocity profile at the inlet.  A resistance and distal pressure boundary condition is imposed at each outlet to achieve the desired flow split $\phi = Q_\text{outlet}/Q_\text{inlet}$, as illustrated in \Cref{fig:junction_schematic}.  The flow split and outlet boundary conditions are related by the equation  
\begin{equation}
    P_\text{dist 1} + \phi R_\text{dist 1} Q_\text{inlet} = P_\text{dist 2} + (1-\phi) R_\text{dist 2} Q_\text{inlet}.
    \label{eq:junction_pressure_balance}
\end{equation}
For simplicity, we choose $P_\text{dist 1} = P_\text{dist 2} = 0 \;\text{Ba}$, and $R_\text{dist 2} = 10^5 \frac{\text{Ba s}}{\text {cm}^3}$, such that, given a desired $\phi$, 
\begin{equation}
    R_\text{dist,1} = \frac{1-\phi}{\phi} \; 10^5 \frac{\text{Ba s}}{\text {cm}^3}.
\end{equation}
\begin{wrapfigure}[17]{r}{0.5\linewidth}
    \includegraphics[width=0.9\linewidth]{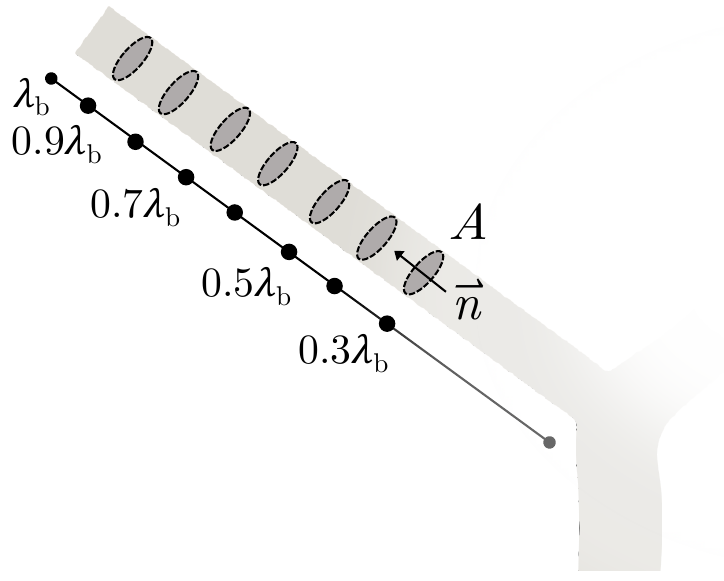}
    \caption{Data flow schematic for prediction of RRI coefficients from bifurcation geometry.}
    \label{fig:centerline_integration}
\end{wrapfigure}
When the simulation is complete, the 3D velocity, $\vec{u}$, and pressure, $p$, results must be converted into 0D quantities, bulk flow $Q$ and average pressure $P$, at the inlet and outlet points.  This is done by integration over the cross-sectional area including the point of interest and normal to the centerline tangent vector (see \Cref{fig:centerline_integration}):  
\begin{equation}
    P = \frac{\int_A p \; dA}{\int_A  dA}, \qquad Q = \int_A \vec{u}\cdot \vec{n} \; dA.
\end{equation}

The bifurcation inlet point is defined by SimVascular's centerline extraction algorithm.  For each bifurcation outlet, we consider seven outlet points (30--90\% of the total branch length, $\lambda_b$), such that we extract 14 inlet-outlet pairs with varying dimensionless outlet lengths, $\lambda$, from a single bifurcation.  We choose a maximum $\lambda_\text{max} = 0.9 \lambda_b$ because we observe the flow to generally have returned to a fully-developed profile, free from the bifurcation effects at this point.  For lengths beyond this, we use existing models for pressure losses in laminar, fully developed pipe flow, i.e. \Cref{eq:poiseuille_mod}. 

\subsubsection{Coefficient Extraction}
We now have a cohort of data points corresponding to inlet-outlet pairs for which we have an inlet flow, $Q_\text{inlet}(t)$, inlet pressure $P_\text{inlet}(t)$, and outlet pressure $Q_\text{outlet}(t)$ (all functions of time). We can then compute $\Delta P(t) = P_\text{inlet}(t_i) - P_\text{outlet}(t_i)$ and find the flow time-derivative using central differences,
\begin{equation}
    \dot{Q}(t) = \frac{Q_{t+1} - Q_{t-1}}{2\Delta t}.
\end{equation}

We find the true RRI coefficients of the bifurcation by finding the least-squares solution to the following system of equations:
\begin{equation}
    \Delta P(t) = R_{\text{lin}} Q(t) + R_{\text{quad}} Q(t)^2 + L \dot{Q}(t),   \; \forall \; t \in \text{simulation timesteps.}
    \label{eq:rri_fit}
\end{equation}
Similarly, the RI coefficients are given by the least-squares solution to 
\begin{equation}
    \Delta P(t) = R_{\text{lin}} Q(t) + L \dot{Q}(t),   \; \forall \; t \in \text{simulation timesteps.}
    \label{eq:ri_fit}
\end{equation}
From these coefficients, we construct the output vector,
\begin{equation*}
    \mathcal{O} = \begin{bmatrix}
    R_\text{lin}, R_\text{quad}, L
    \end{bmatrix},
\end{equation*}
the dimensionless output vector, as defined in \Cref{eq:rri_nd},
\begin{equation*}
    \mathcal{O}^* = \begin{bmatrix}
    R_\text{lin}^* = R_\text{lin}\frac{Q_c}{P_c}, \;\;\;\; R_\text{quad}^*=R_\text{quad}\frac{Q_c^2}{P_c},\;\;\;\;  L^* = L \frac{t_c}{Q_c}
    \end{bmatrix},
\end{equation*}
and the z-normalized, dimensionless output vector, 
\begin{equation}
    \mathcal{O}_{z,i}^* = \frac{(\mathcal{O}^*_i - \mu_i)}{\sigma_i}
\end{equation}
which is used as the ground truth in training the neural network.  Again, $\mu_i$ and $\sigma_i$ are the mean and standard deviation of the distribution of dimensionless RRI and RI coefficients in the training set. 

\subsection{RRI Model Deployment in Trees}
\begin{figure}[htbp]
    \centering
    \includegraphics[width=0.95\linewidth]{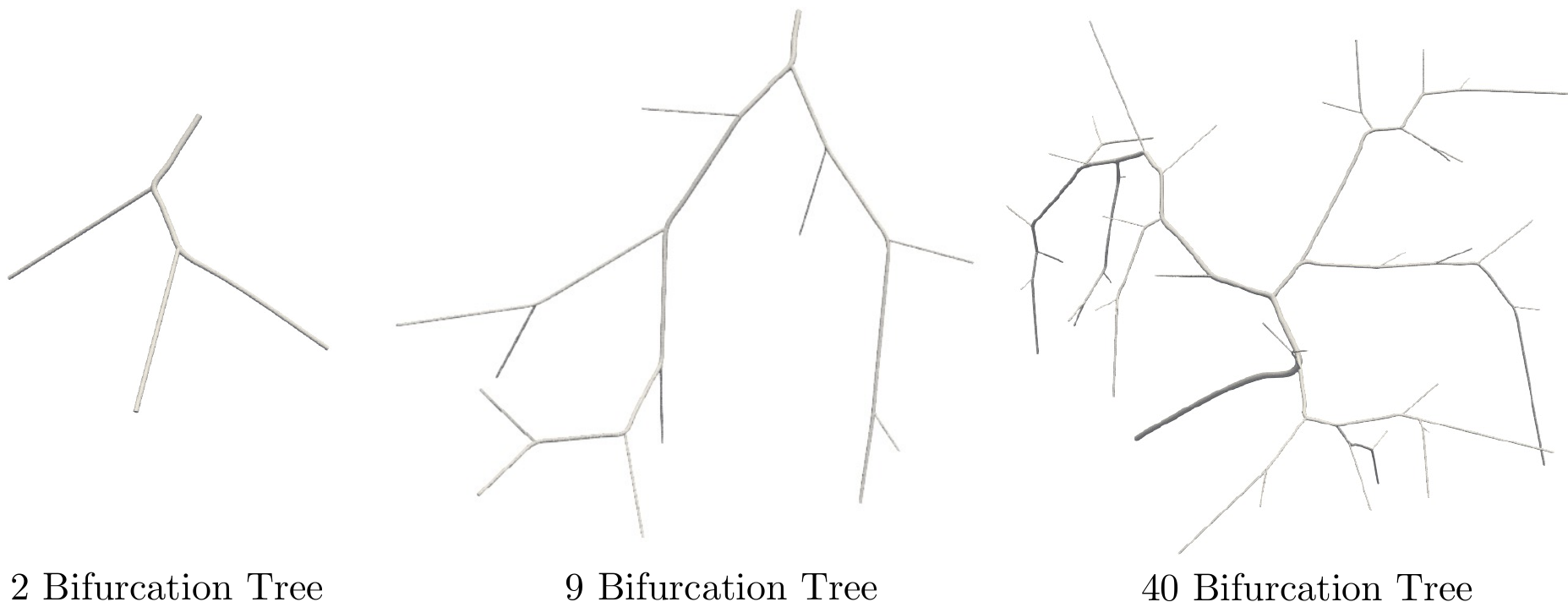}
    \caption{Vascular trees used to analyze the efficacy of RRI junction handling.}
    \label{fig:cco_trees}
\end{figure}

We next examine the effect of the RRI model on 0D ROM simulation accuracy for vascular trees of varying sizes, shown in \Cref{fig:cco_trees}.  The tree geometries are generated using \texttt{svVascularize} \cite{Sexton2025RapidBiomanufacturing}.  These trees have highly regular geometries, featuring idealized cylindrical branches consistent with the assumptions used in deriving the 0D model.  The absence of error introduced by the branch geometry allows us to better isolate the effects of the bifurcations.  As with the synthetic junctions, we then use SimVascular's Python API to generate surface meshes and \texttt{svMultiPhysics} for 3D flow simulations.  We prescribe plug flow at the inlet and resistance boundary conditions at the outlets.  Again, we apply inlet Reynolds numbers ranging from 0 to 5,500 to capture the full range of observed physiologic Reynolds numbers.

\subsubsection{Standard 0D Pipeline}
To run 0D simulations with standard bifurcation handling, we first use SimVascular's ROM simulation tool \footnote{\url{https://github.com/SimVascular/svMultiPhysics}}  to convert the 3D geometry into the standard 0D description of its electric circuit analog.  This produces a 0D input file containing a list of the vessels, described by circuit elements with characteristic values computed according to \cref{eq:characteristic_values}, and a list of junctions describing the connectivity of the vessels.  The input file also contains information about the bulk vascular geometry, i.e., vessel radii, lengths, and tangents.  The standard 0D input file does not contain any bifurcation-specific circuit elements.

The system of equations governing the circuit described by the standard input file is solved with SimVascular's \texttt{svZeroDSolver} \footnote{\url{https://github.com/SimVascular/svZeroDSolver}} \cite{Menon2025SvZeroDSolver:Simulations}.  This application uses a generalized-alpha time advancement scheme and a Newton-Raphson solver to find the optimal solution at each timestep.

\subsubsection{RRI 0D Pipeline}
To run a 0D simulation using RRI bifurcation handling, we start by modifying the standard input file.  First, we remove the circuit elements describing the vessels, as the vessels will now be handled as part of the junction outlets.  Next, we determine RRI coefficients $\mathcal{O} = [R_\text{lin}, R_\text{quad}, L]$ for each junction outlet, following the procedure shown in \Cref{fig:geo_to_rri}.  

The standard 0D input file contains geometric information needed to compute $\alpha$, $\lambda$, and $\theta$ for each bifurcation outlet.  We approximate the flow ratio $\phi$ based on the circuitry downstream of each outlet.  Rearranging \Cref{eq:junction_pressure_balance}, and given 0 Ba distal pressure boundary conditions, we have
\begin{equation}
   \phi = \frac{R_\text{2}}{(R_\text{1}+R_\text{2})}.
   \label{eq:flow_split}
\end{equation}
We find $R_\text{dist,1}$ and $R_\text{dist,2}$ using the recursive rule:
\begin{equation}
  R=\begin{cases}
    R_\text{outlet} + (R_\text{dist,1}^{-1}+R_\text{dist,2}^{-1})^{-1}, & \text{if outlet ends in another bifurcation}.\\
    R_\text{outlet} + R_\text{dist,BC}, & \text{if outlet ends in a boundary condition (BC)}.
  \end{cases}
  \label{eq:resistance_recursion}
\end{equation}
We compare the flow splits found using this technique to those observed in 3D simulation and include the results in \Cref{app:flow_splits}.  We find good agreement and invariance to the flow magnitude.

Having determined the geometric input feature vector $\mathcal{G}^*$, we verify that the neural network inputs and outputs are within the ranges on which the network was trained.  If a bifurcation length is outside the range of those in the training set, we make a correction to $R_\text{lin}$ and $L$.  We first determine the amount by which the length exceeds the training set range,
\begin{equation}
    l_\text{add} = \min(0,  \lambda - \lambda_\text{min})l_c + \max(0, \lambda - \lambda_\text{max})l_c,
\end{equation}
and then add the corresponding resistance and inductance, 
\begin{equation}
R_\text{lin, mod} = R_\text{lin} + l_\text{add} \frac{8 \pi \mu}{A_\text{outlet}^2}, \qquad L_\text{mod} = L + l_\text{add} \frac{\rho}{A_\text{outlet}}.
\label{eq:poiseuille_mod}
\end{equation}
We create modified RRI and RI 0D input files containing resistance/inductance-free vessels and junctions containing RRI/RI coefficients computed for each bifurcation outlet.  

\subsubsection{Numerical solution strategy}
The introduction of large quadratic resistance terms fundamentally alters the mathematical structure of the circuit equations, injecting major nonlinearities into the system with significant implications for numerical solution strategies. The resulting system exhibits several computational challenges: (1) the Jacobian matrix becomes highly flow-dependent, degrading the convergence of standard Newton-Raphson methods, and (2) quadratic terms with opposing signs can create ill-conditioned systems or non-physical solutions.
Consider the extreme case where a bifurcation contains only quadratic resistors of opposite signs. The pressure balance equation becomes:
\begin{equation}
    R_\text{quad,1}Q_1^2 = R_\text{quad,2}Q_2^2, \; \text{ where } \; R_\text{quad,2} < 0 < R_\text{quad,1},
    \label{eq:ill_posed_case}
\end{equation}
which admits no physical solution except the trivial case $Q_1 = Q_2 = 0$. This mathematical singularity reflects the fundamental challenge of capturing complex bifurcation physics with simplified circuit models.  We note that stenosis resistors described in \Cref{eq:standard_mass_conservation} and \Cref{eq:standard_mom_conservation} have been used in standard 0D frameworks without numerical complications likely because (1) circuits based on physiological anatomies generally do not include as many stenosis resistors of the magnitude we are considering, and (2) stenosis is always positive, due to the absolute value in its formulation, preventing the ill-posed scenario in \Cref{eq:ill_posed_case}.

To address these numerical challenges, we reformulate the problem as a constrained optimization problem rather than a root-finding problem. This approach offers several algorithmic advantages: (1) objective function minimization is more robust to local minima than exact root finding, (2) constraint handling naturally incorporates physical conservation laws while allowing for uncertainty in the predicted RRI or RI coefficients and predicted flow splits, and (3) optimization solvers provide better convergence guarantees for nonlinear systems.

The optimization formulation minimizes deviation from the RRI or RI model and predicted flow split at bifurcations.  To keep the two terms at comparable sizes, pressure contributions are scaled by the square of the prescribed inlet flow, while flow contributions are scaled by the inlet flow:
\begin{equation}
\text{minimize} \quad Z \; = \sum_\text{bifurcation outlets} \left( 
\frac{\overbrace{P_\text{in} - P_\text{out} - (R_{\text{lin}} Q + R_{\text{quad}} Q^2 + L \dot{Q})}
^{\text{junction pressure equation (RRI or RI model)}}}{Q_\text{inflow BC}^2} \right)^2 + \left( \frac{\overbrace{(\phi Q_\text{in} - Q_\text{out})}^{\text{flow split}}}{Q_\text{inflow BC}} \right)^2
\end{equation}
The optimization is subject to exact enforcement of mass conservation, boundary conditions, and temporal evolution constraints:
\begin{align}
Q_\text{in} - Q_\text{out,1} - Q_\text{out,2} &= 0 && \forall \text{ junctions} && \text{(mass conservation)} \label{eq:mass_conservation_junction}\\
Q_\text{in} - Q_\text{out} &= 0 && \forall \text{ vessels} && \text{(mass conservation)} \label{eq:mass_conservation_vessel}\\
P_\text{in} - P_\text{out} &= 0 && \forall \text{ vessels} && \text{(pressure continuity)} \label{eq:pressure_continuity_opt}\\
Q_\text{in,inlet} - Q_\text{inflow BC} &= 0 && && \text{(inlet flow boundary condition)} \label{eq:inlet_bc}\\
P_\text{out} - R_{\text{dist,BC}} Q_\text{out} &= 0 && \forall \text{ outlet vessels} && \text{(outlet resistance boundary conditions)} \label{eq:outlet_bc}\\
\dot{P} = \dot{Q} &= 0 && && \text{(steady state)} \label{eq:steady_state}\\
\dot{P} - \frac{P_t - P_{t-1}}{\Delta t}  &= 0 && &&\text{(transient)} \label{eq:transient}\\
\dot{Q} - \frac{Q_t - Q_{t-1}}{\Delta t} &= 0&& && 
\end{align}

Since the mass conservation equations introduce no error and must be enforced, we include them as constraints.  We also include a constraint to enforce continuity of pressure over vessels, which now act as wires connecting junctions to each other.  For steady simulations, we add the constraint that the solution time derivative is zero.  For transient simulations, we define the solution time derivative using backward differences, as in \texttt{svZeroDSolver}.  Finally, we impose the inlet flow boundary condition and outlet resistance boundary conditions as constraints.

We construct a symbolic representation of the problem using CasADi, which uses algorithmic differentiation to provide the gradients and Hessians needed to solve the problem with Ipopt, an interior point solver \cite{Andersson2019CasADi:Control, Wachter2006OnProgramming}.

\section{Results}

\begin{figure}[htbp]
    \centering
    \includegraphics[width=0.99\linewidth]{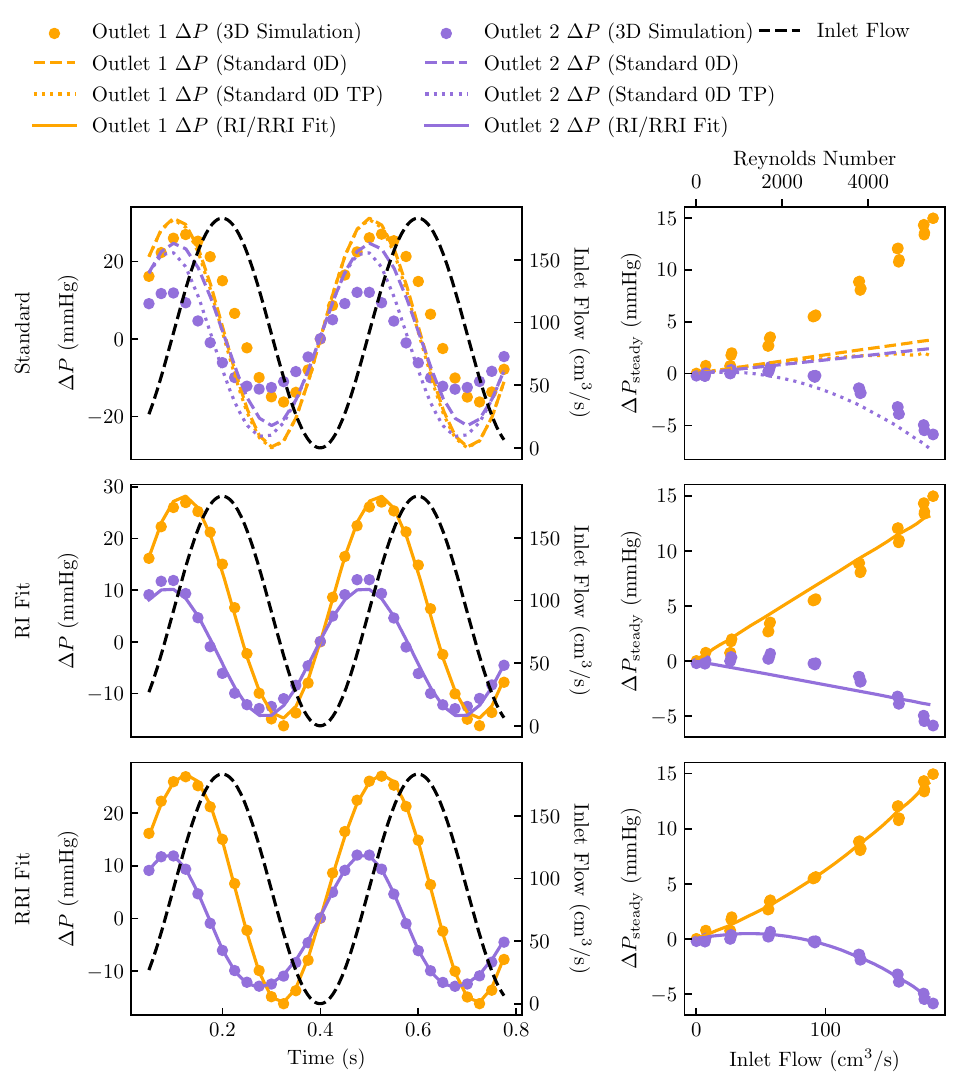}
    \caption{RRI (top) and RI (middle) fits to flow and pressure drop data from 3D simulation of an isolated bifurcation.  Standard 0D pressure drop predictions over the portion of the outlet branch included in the bifurcation (bottom), assuming constant pressure and conservation of total pressure (TP). Transient flow and pressure profiles (left) and steady component of $\Delta P$ (right), computed as $\Delta P_\text{steady} = \Delta P - L \dot{Q}$.}
    \label{fig:isolated_fit}
\end{figure}

For each bifurcation inlet-outlet pair in the cohort of synthetically-generated isolated bifurcations, we fit RRI and RI coefficients from the bulk flow and pressure extracted from transient simulations, according to \Cref{eq:rri_fit} and \Cref{eq:ri_fit}, respectively.  In a representative example shown in \Cref{fig:isolated_fit}, we saw that the RRI fit matched the data slightly better than the RI fit.  In particular, the inclusion of a quadratic resistor allowed for better capturing of the relationship between steady pressure drop and flow, shown in the right panel.

Across the entire cohort of $\approx$4,000 pairs, the RRI fits had an average $R^2_\text{RRI} = 0.9998$, while the RI fits had an average least-squares residual of $R^2_\text{RI} = 0.9788$.  

We also plot the pressure drop that would be predicted by the standard 0D model over the portion of the branch included in the bifurcation using the resistance and inductance determined by \Cref{eq:physics_based_char_vals} and either a pressure continuity (most commonly used) or conservation of total pressure assumption (used in some 1D models) at the bifurcation.  That is,
\begin{align}
    \Delta P_\text{pressure continuity}  &= \frac{8 \pi \mu l}{A^2}Q + \frac{\rho l}{A}\dot{Q},  
    \label{eq:pressure_continuity}
\intertext{and}
    \Delta P_\text{total pressure conservation}  &= \frac{8 \pi \mu l}{A^2}Q + \frac{\rho l}{A}\dot{Q} + \frac{\rho}{2}(U_\text{outlet}^2 - U_\text{inlet}^2).
    \label{eq:total_pressure_cons}
\end{align}
The standard 0D model greatly overestimated the inductances while significantly underestimating the resistances.  Enforcing conservation of total pressure somewhat reduced the error in the steady component, $\Delta P_\text{steady}$, thereby also improving the transient prediction slightly.  Even using conservation of total pressure, however, there was still a significant error in the prediction of $\Delta P_\text{steady}$, particularly at higher Reynolds numbers.

\begin{figure}[htbp]
    \centering
    \includegraphics[width=1.0\linewidth]{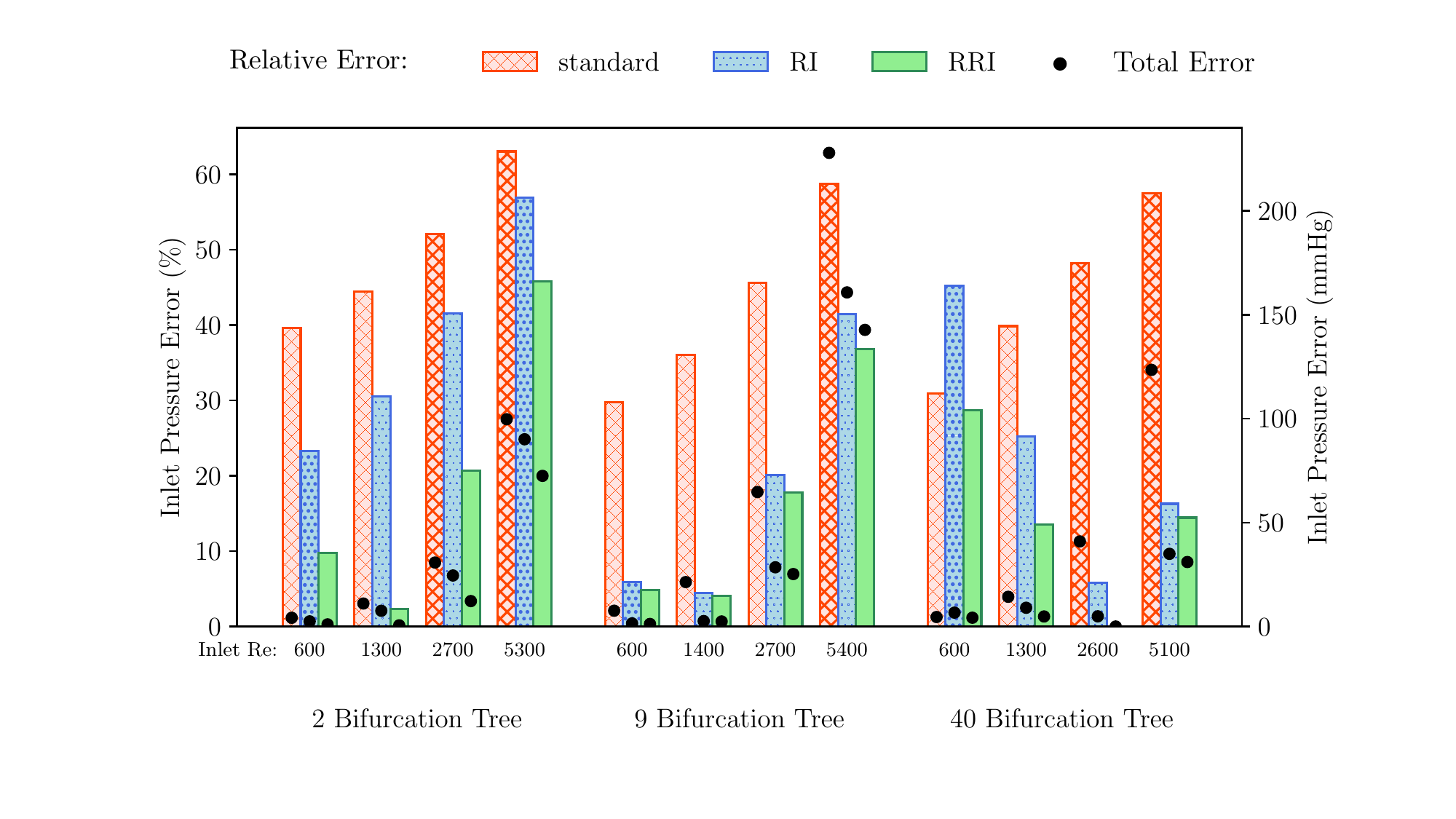}
    \caption{Inlet relative pressure error, defined by comparison to 3D simulations for standard and RRI bifurcation handling for three vascular trees of increasing size, and four different inlet Reynolds numbers. Black circles mark the absolute inlet pressure error.}
    \label{fig:steady_errors}
\end{figure}

\begin{figure}[htbp]
    \centering
    \includegraphics[width=0.99\linewidth]{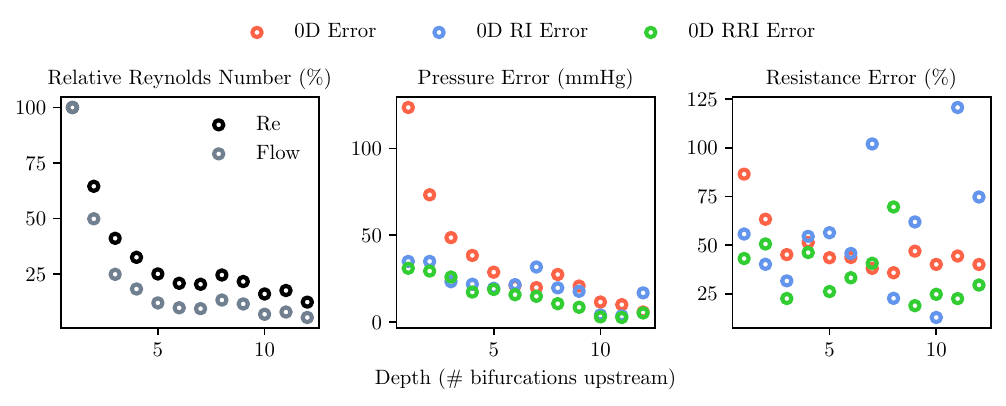}
    \caption{Bifurcation statistics for bifurcations at various depths of the 40 bifurcation tree, where depth is defined as the total number of bifurcations downstream of the bifurcation in question.  Left: Flow and Reynolds number at each junction inlet, relative to the flow and Reynolds number at the tree inlet.  Center: pressure error at maximum inflow.  Right: resistance error over each junction, for the largest inlet flow value.}
    \label{fig:depth_plots}
\end{figure}

We ran 3D and 0D (with the standard, RI, and RRI bifurcation handling methods) steady simulations for three vascular trees of varying depth, varying the Reynolds number at the inlet of the tree.  We chose inlet Reynolds numbers ranging from $\approx$600 to $\approx$5,500. (Notably, Reynolds number quickly decays down the trees (\Cref{fig:depth_plots}), so we encounter a wide range of Reynolds numbers across the tree.)  We defined an error for each 0D simulation by comparison to the 3D simulation (\Cref{fig:steady_errors}).  The RRI and RI bifurcation handling yielded more accurate results than the standard bifurcation handling.  Both the absolute and relative errors for the standard 0D model increased with Reynolds number.  While the RI model generally outperformed the standard 0D model, for the 40 junction tree, it yielded a higher error than the 0D model at the lowest Reynolds number. Notably, though, this corresponded to a low absolute error, and in the higher Reynolds number regime, the RI model outperformed the standard 0D model by a large absolute error.  The RRI model generally outperformed the RI model, and outperforms the standard 0D model across the full range of Reynolds numbers.  The RRI model makes a bigger improvement over standard bifurcation handling for larger trees.  Across all trees and Reynolds numbers, we observed an average error of 55 mmHg (45\%) for the standard bifurcation handling, 31 mmHg (26\%) for the RI bifurcation handling, and 25 mmHg (17\%) for the RRI bifurcation handling.

\begin{figure}[htbp]
    \centering
    \includegraphics[width=0.99\linewidth]{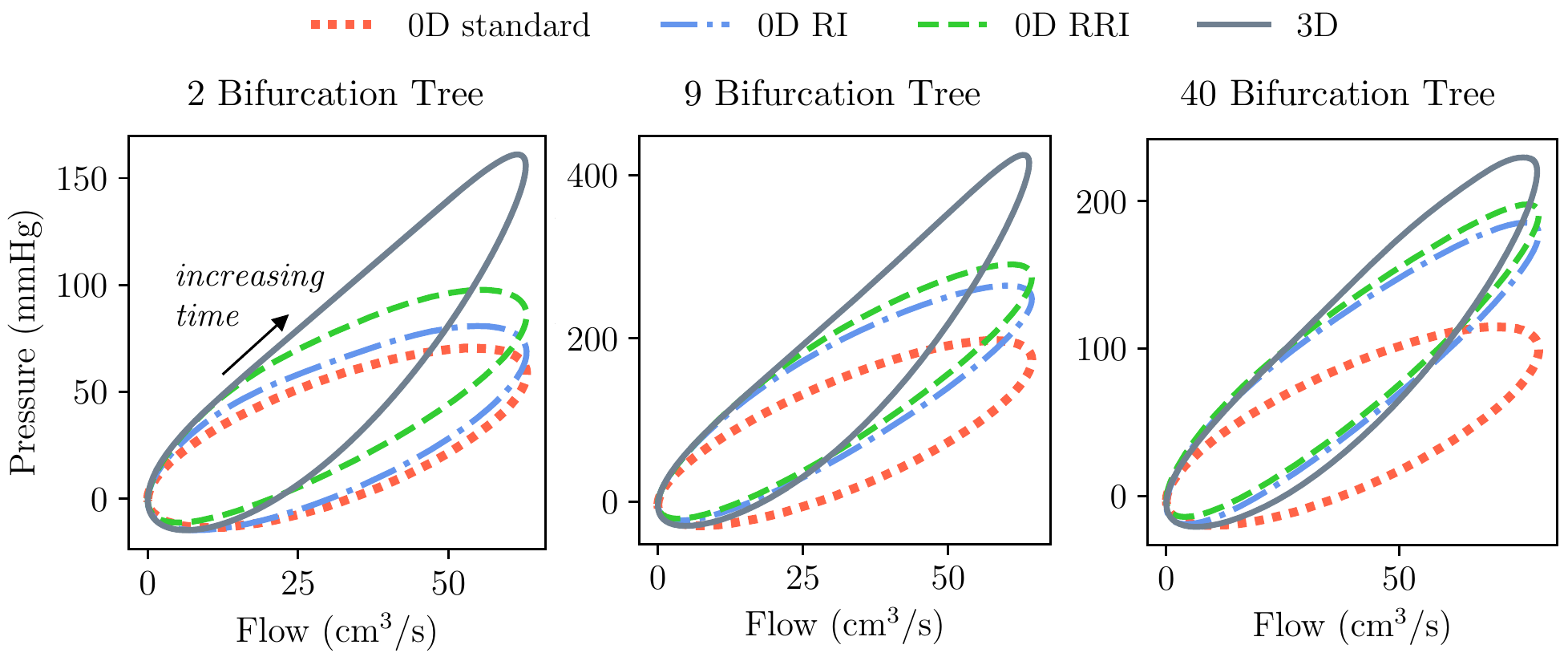}
    \caption{Inlet pressure-flow curves for transient simulations for three vascular trees.  Solutions from the standard 0D model, 0D RI model, and 0D RRI model are shown in comparison to the 3D simulation.}
    \label{fig:unsteady_plots}
\end{figure}

We ran transient simulations where the inlet flow varied in time, following a sinusoidal profile with a period of 0.4 seconds, again chosen to be representative of the systolic portion of the cardiac cycle in high-Reynolds number vasculatures.  The inlet Reynolds number varied from 0 to $\approx$5,500.  We show inlet pressure-flow curves displaying the 3D solution, standard 0D solution, RRI 0D solution, and RI 0D solution in \Cref{fig:unsteady_plots}.  Again, in all three trees, the RRI and RI solutions tracked the 3D simulation solution better than the standard 0D model.

\begin{figure}[htbp]
    \centering
    \includegraphics[width=0.99\linewidth]{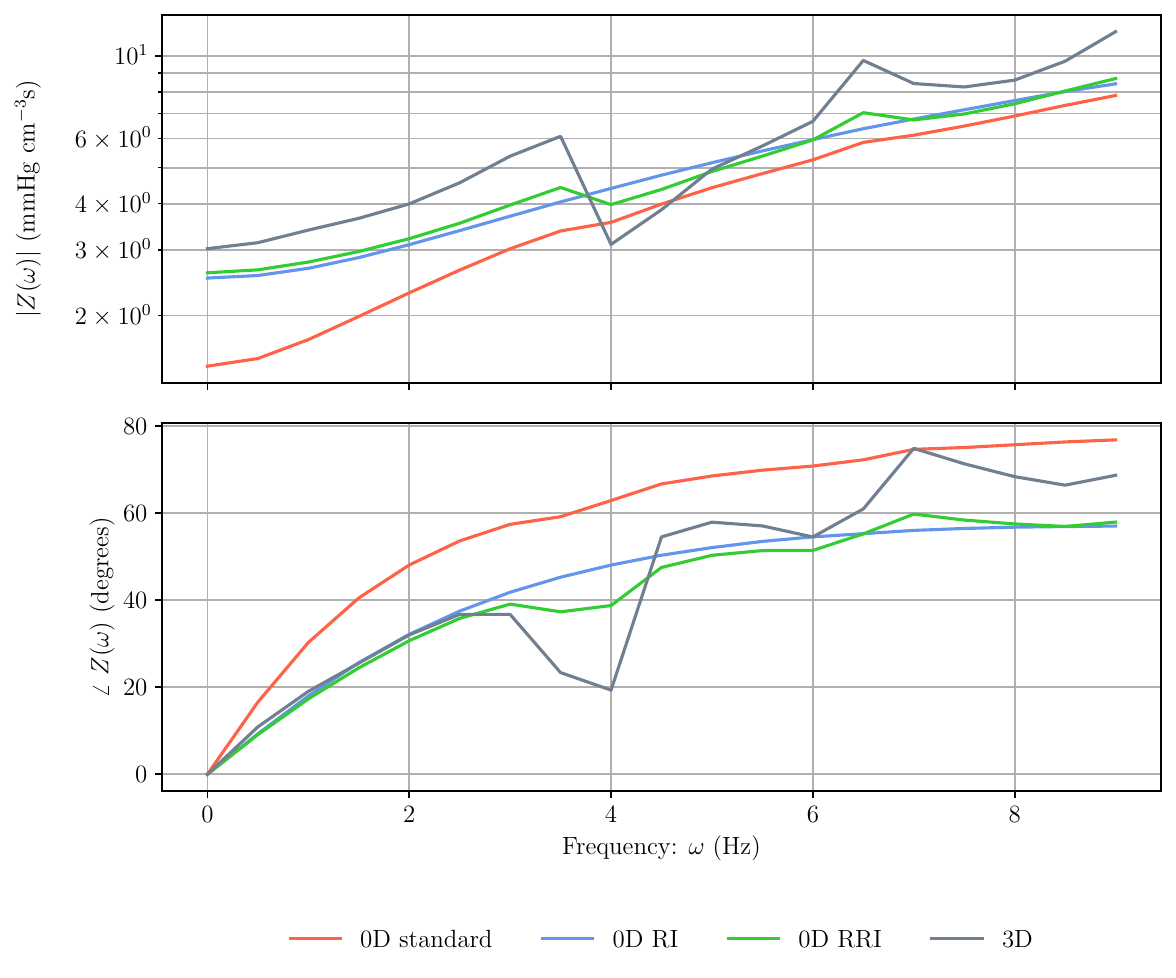}
    \caption{Impedance frequency response for the 40 outlet tree.  Impedance magnitude is shown in the upper panel, and phase is shown in the lower panel.}
    \label{fig:impedance}
\end{figure}

For the largest tree, we computed the input impedance as predicted by 3D, standard 0D, RI 0D, and RRI 0D simulations, shown in \Cref{fig:impedance}.  To compute impedance, we applied a realistic inflow waveform based on a patient-specific aortic model \cite{LaDisa2025Vascular0011_H_AO_H, Jr.LaDisa2011ComputationalAnastomosis, Wilson2013TheResults}.  Since the inlet diameter of the tree was different from that of the patient-specific aorta, we scaled the flow such that the maximum Reynolds number was 5,500, as observed physiologically.  The impedance was then computed from Fourier transforms of the imposed flow and resulting pressure signals.
\begin{equation}
    Z(\omega) = \frac{\Delta \hat{P}(\omega)}{\hat{Q}(\omega)}  =\frac{\mathcal{F}(\Delta P (t))}{\mathcal{F}(Q(t))}.
\end{equation}

The RRI 0D model again matched the 3D solution best, followed by the RI 0D model and the standard 0D model.  We note that since our trees only included resistors and inductors, the frequency response looked different from those that include capacitors \cite{Qureshi2018CharacteristicApproach, Olufsen1999StructuredArteries, Taylor1966WaveTubes}.  While capacitors are associated with a phase-lag (negative phase) and damping effects (decreasing magnitude), inductors generally cause a phase-lead (positive phase) and amplification of oscillations (increasing magnitude).

For the largest tree, we analyzed each bifurcation independently, observing trends relating to the bifurcation's depth in the tree, which we defined as the number of bifurcations upstream of it.  We plot the flow and Reynolds number at each bifurcation inlet (normalized by the inlet flow and Reynolds number) in the left panel of \Cref{fig:depth_plots}.  The two quantities followed similar profiles, decreasing with bifurcation depth in a similar profile to that of the standard 0D pressure error in the center panel.

In the center panel of \Cref{fig:depth_plots}, we plot the pressure error from the largest Reynolds number steady simulation at each bifurcation inlet against its depth.  For deep bifurcations, all three models had similar errors.  Close to the inlet, however, the error of the standard bifurcation handling method increased significantly, with the largest pressure error occurring at the bifurcations closest to the tree inlet.  The RRI and RI models maintained a relatively constant error across the tree, although it decreased slightly with bifurcation depth.

We computed the resistance at each bifurcation at the highest inlet Reynolds number as predicted by the 0D model with standard, RI, and RRI bifurcation handling. We plot the error compared to the 3D simulation in the right panel of \Cref{fig:depth_plots}. The 0D model with RRI bifurcation handling generally made a lower error than the standard 0D model across the range of depths.  The RI 0D model improved on the standard 0D model for the less deep (proximal) bifurcations, but made less accurate estimates of resistance for the more distal bifurcations.

In this work, to fully capture the bifurcation effects, we define the bifurcation outlets (along the daughter branches) to be more distal to the bifurcation than in the widely used method of branch-junction distinction proposed in \cite{Pfaller2022AutomatedFlow}, where the bifurcation ends as soon as the outlet cross-sectional areas no longer intersect.  We call our definition ``partial-branch'' and the definition in \cite{Pfaller2022AutomatedFlow} ``no-branch''.  In most cases, however, there remains a section of branch between the outlet of the previous bifurcation and the inlet of the next bifurcation (or the end of the branch), which we account for using the standard 0D Poiseuille resistance and inductance, \Cref{eq:poiseuille_mod}.  Since the distinction between bifurcation and branch is arbitrary, though, we may also consider a third bifurcation definition where the bifurcations include the entirety of outlet branches and the 0D Poiseuille resistance is not used at all. We call this definition ``full-branch''. We perform an additional study to determine the maximum accuracy achievable using each bifurcation definition.  

For each of these bifurcation definitions, we fit RRI and RI resistances (we neglect inductance for this steady analysis) to each bifurcation in each of the three trees using the four steady simulations at four different inlet Reynolds numbers,
\begin{align}
    \Delta P_\text{Re} &= R_{\text{lin,RRI}}\, Q_\text{Re} + R_{\text{quad}} Q_\text{Re}^2 && \forall \text{ Re} \in [600, 1300, 2700, 5500] \quad \text{(RRI)}, \\
    \Delta P_\text{Re} &= R_{\text{lin,RI}}\, Q_\text{Re} && \forall \text{ Re} \in [600, 1300, 2700, 5500] \quad \text{(RI)}.
\end{align}
For each tree, we then run steady 0D simulations at the four steady Reynolds numbers considered in previous analyses using the three bifurcation definitions and the fitted coefficients.  We again determine the accuracy of each simulation by comparison to the 3D solution and report the relative inlet pressure errors for each tree (averaged over the four simulations of varying inlet Reynolds number) in \Cref{tab:coverage_errors}.  Generally, the full-branch bifurcation definition achieved the lowest errors, followed by the partial-branch definition.  The no-branch definition yielded significantly higher errors than the other two.  The RRI fits were substantially more accurate than the RI fits.

\begin{table}[htpb]
\begin{center}
\begin{tabular}{p{1.5in} p{0.31in} p{0.5in} p{0.31in} p{0.5in} p{0.31in} p{0.5in}}

Bifurcation definition: &\multicolumn{2}{c}{No-Branch} & \multicolumn{2}{c}{Partial-Branch} & \multicolumn{2}{c}{Full-Branch}\\
Fitting method: & RI & RRI  & RI & RRI  & RI & RRI \\
    \cmidrule(lr){2-3}\cmidrule(lr){4-5}\cmidrule(lr){6-7}

\multicolumn{7}{l}{\textit{Relative Error}}\\
2 Bifurcation Tree & 0.71 & 0.62 & 0.20 & 0.17 & 0.32 & 0.08 \\
9 Bifurcation Tree & 0.48 & 0.45 & 0.30 & 0.08 & 0.40 & 0.05 \\
40 Bifurcation Tree & 0.47 & 0.44 & 0.33 & 0.04 & 0.35 & 0.05 \\
\\
\multicolumn{7}{l}{\textit{Total Error (mmHg)}}\\
2 Bifurcation Tree & 48.35 & 47.52 & 12.37 & 11.97 & 11.81 & 4.88 \\
9 Bifurcation Tree & 88.84 & 87.46 & 29.54 & 10.17 & 28.23 & 7.44 \\
40 Bifurcation Tree & 48.61 & 47.69 & 13.12 & 4.27 & 12.96 & 6.21 \\

\end{tabular}
\caption{Relative and total inlet pressure error, averaged over four steady simulations with inlet Reynolds numbers varying from $\approx$ 600 to 5,500 for different bifurcation definitions.}
\label{tab:coverage_errors}
\end{center}
\end{table}

\section{Discussion}

We found that the standard 0D model made a large error in predicting pressure drops over isolated bifurcations, which in turn caused significant pressure error in 0D models of vascular trees that was substantially reduced by incorporating the RRI or RI bifurcation models. The large discrepancy between pressure drops predicted by existing bifurcation handling techniques and the 3D solution, shown in the bottom panel of \Cref{fig:isolated_fit}, emphasizes the need for a more sophisticated bifurcation handling technique.  Although total pressure conservation yielded slightly better results than continuity of pressure by more accurately capturing the steady contribution to the pressure drop, neither approach was able to capture the steady or transient flow behavior in the bifurcation fully. In the steady, transient, and frequency response analyses of vascular trees, the large error between the standard 0D and 3D simulation results was again apparent.  Notable trends observed in both isolated and in-tree studies were overestimation of inductance and underestimation of steady-state resistance, particularly at high Reynolds numbers.  In both isolated (\Cref{fig:isolated_fit}, bottom right panel) and in-tree (\Cref{fig:steady_errors}) analysis, both absolute and relative errors for the standard 0D model increased with Reynolds number, suggesting the presence of steady, non-linear effects that were not captured by the standard 0D model and are more pronounced at higher Reynolds numbers, in turn supporting the use of the quadratic resistor to capture this behavior.

In contrast, the RRI and RI bifurcation methods were able to more closely replicate 3D simulation.  In the isolated bifurcation study, RRI and RI models captured the flow-pressure relationship extremely closely (\Cref{fig:isolated_fit}).  The transient flow-pressure curve typically forms a closed loop, as in \Cref{fig:unsteady_plots}; the collapse of the loop onto a one-to-one steady flow-pressure curve, as in the right column of \Cref{fig:isolated_fit} upon subtraction of the inductor contribution, $L \dot{Q}$, indicates that the inductor fully captured the unsteady flow behavior in the bifurcation.  Moreover, in the upper-right corner of \Cref{fig:isolated_fit}, we see that the linear and quadratic resistors described the steady flow-pressure relationship extremely well. The RI $Q-\Delta P_\text{steady}$ plot revealed that the linear resistance alone was not able to fully capture the relationship between flow and pressure loss, suggesting that the RI model may not be able to achieve high accuracy across all Reynolds numbers.  Overall, however, while clearly not as accurate as the RRI fit, the RI fit represented 3D flow behavior in bifurcations quite well.

When the RRI and RI methods were incorporated into 0D vascular tree models, the RRI bifurcation handling again yielded results closest to the 3D simulation.  The RI bifurcation method was only slightly less accurate, and both models outperformed the standard 0D model substantially (\Cref{fig:steady_errors,fig:unsteady_plots,fig:impedance}). While the RI model does not perform as robustly across varying Reynolds numbers as the RRI model, it may be a good alternative to the RRI model as it avoids the practical complexities introduced by the quadratic term, without sacrificing much accuracy.  Unsurprisingly, the RI and RRI models make the biggest improvements over the standard 0D model for larger trees that include more bifurcations, as more bifurcations present more opportunities for the introduction of significant error into the standard 0D model.  

Our analysis of each bifurcation in the largest tree yielded several insights into the flow and pressure behavior at different tree depths.  The left panel of \Cref{fig:depth_plots} shows the simultaneous decay of flow and Reynolds number across the tree.  It is obvious that flow must decrease at each bifurcation as it is divided between the two outlets; it is notable that the Reynolds number also decreased very similarly.  The daughter radii in a vessel bifurcation are generally determined by Murray's law, 
\begin{equation}
    R_\text{inlet}^3 = R_\text{outlet,1}^3 + R_\text{outlet,2}^3,
\end{equation}
such that the total outlet area is larger than the inlet area, causing a reduction in both velocity and vessel diameter that both contribute to a decrease in Reynolds number.  As observed in the steady analyses above, Reynolds number determines the magnitude of non-linear effects in the bifurcation, and therefore the degree to which outlet resistance differs from the standard 0D model's prediction. Meanwhile, flow scales the resistance error, such that high flows amplify errors in the 0D model's estimate of resistance while low flows mute these errors.

The trends in flow and Reynolds number explain much of the pressure and resistance errors observed across the tree.  As expected, the standard 0D model made larger errors in resistance at more proximal bifurcations, where the Reynolds number was higher. Since the flow was largest in the proximal bifurcations, this error was amplified, producing a very high pressure error.  In contrast, the RRI model yielded lower resistance errors which were not as dependent on Reynolds number, in turn producing lower pressure errors.  The RI model made low resistance errors on the high Reynolds number proximal bifurcations, but was less accurate on the low Reynolds number bifurcations.  This echoes the earlier observation that a linear resistance alone cannot achieve high accuracy at all Reynolds numbers.  Since the low Reynolds number bifurcations are also in low-flow regimes, however, these resistance errors are not very impactful, and the RI model is able to achieve low pressure errors similar to those of the RRI model.

In our study of three bifurcation definitions, which included the downstream branches in the bifurcation to varying extents, we found that when more of the downstream branch was included in the junction, we generally achieved higher accuracy.  This was expected because as we included more of the downstream branches in bifurcations, we fit a greater fraction of the vasculature to the 3D result, and used the 0D Poiseuille resistances for a smaller fraction. The RRI fitting method clearly outperformed the RI method, justifying the use of the quadratic resistor to achieve higher ROM accuracy.

Notably, the no-branch bifurcation definition yielded a much higher error than the partial-branch definition.  This is explained by the fact that the non-linear flow effects caused by the bifurcation extend beyond the split, and the flow only resumes a fully-developed, laminar profile that can be represented by the 0D Poiseuille resistances some distance down the outlet branch.  We also observed this in the 3D bifurcation flow and pressure fields shown in \Cref{app:3d_fields}.  The reduced error of the partial-branch versus no-branch bifurcation definition supports our decision to use the partial branch approach in this work.

We saw an additional, albeit smaller improvement in accuracy from a switch from the partial-branch to full-branch junction definition, suggesting that, even in the fully developed region some distance downstream of the bifurcation, there was some error in the standard 0D Poiseuille resistance estimate.  This motivates future work to predict more accurate 0D coefficients for branches, as well as junctions in 0D modeling frameworks.

\section{Limitations and Future Work}

While this work successfully enhances 0D ROM accuracy, we note several limitations and discuss next steps to overcome them.  First, we tested the RRI and RI models on synthetically generated, idealized rigid-wall trees that are not fully representative of the patient-specific vasculatures for which we ultimately hope to use these techniques.

The synthetic trees have highly regular geometries compared to human vasculature, which may feature curvature, cross-sectional area variation, and diseased sections such as stenoses and aneurysms.  Future work should analyze the RRI and RI models' ability to handle these geometries. This may, in turn, require more detailed geometric representations of bifurcations as input to the machine learning model that predicts the RRI coefficients and more complex machine-learning models, such as graph neural networks.  

Another complication of human vasculature is the presence of junctions with more than two outlets as described in \cite{Pfaller2022AutomatedFlow}.  To handle these cases, future work will need to expand the bifurcation method to predict RRI coefficients for many-outlet junctions or use an alternative junction definition where the many-outlet junctions are split into a series of bifurcations.  

Additionally, to more fully capture the vascular hemodynamics, the compliance of the vessels should be considered.  This may be done by the inclusion of a capacitor in the bifurcation-handling model, such that it consists of a linear resistor, quadratic resistor, inductor, and capacitor (RRIC).  In this case, 3D fluid-structure interaction simulations must be run to generate the data used to train the ML models to predict the RRIC coefficients, introducing complexity and computational cost compared to the rigid-wall simulations used as training data in this work.

A drawback of the RRI model may be the increased complexity of finding the model solution.  The inclusion of the quadratic term introduces non-linearity into the system of equations governing the 0D model that makes it harder for iterative solvers (e.g., Newton-Raphson root finder, used in \cite{Menon2025SvZeroDSolver:Simulations}) to find the zeros of the system's residual, resulting in increased solve times.  Furthermore, in cases where there is no exact solution, it becomes necessary to switch to an optimization framework that minimizes the residual, rather than finding its zero.

Finally, as we see in \Cref{tab:coverage_errors}, ROM error can only be reduced to a certain extent by modeling the bifurcations as we define them in this work.  This error may be further reduced by modeling the entire vasculature, a potential direction of further study, but ultimately will always include some error due to the simplified form of the 0D ROM model.

\section{Conclusions}
This work presents a hybrid numerical framework that combines machine learning with physics-based reduced-order modeling to address fundamental challenges in cardiovascular flow simulation. We develop an enhanced resistor-resistor-inductor (RRI) model and a simplified resistor-inductor (RI) variant that capture complex bifurcation hemodynamics while maintaining the computational efficiency of 0D models.

Our key numerical contributions include: (1) a physics-based non-dimensionalization scheme that reduces training data requirements by representing geometrically similar bifurcations with unified parameters, (2) an optimization-based solution strategy using interior-point methods to handle nonlinear system equations where traditional Newton-Raphson approaches fail, and (3) constraint handling that ensures mass conservation while accommodating machine learning prediction uncertainties.

Comprehensive validation against high-fidelity 3D finite element simulations across Reynolds numbers from 0 to 5500 demonstrates substantial accuracy improvements. The RRI method reduces average pressure errors from 54 mmHg (45\%) for standard 0D models to 25 mmHg (17\%), while the RI variant achieves 31 mmHg (26\%) error. Performance gains are most pronounced at high Reynolds numbers and in extensive vascular networks, where standard approaches exhibit the largest deviations from 3D solutions.

The computational framework maintains real-time simulation capabilities while providing clinically relevant accuracy improvements. This enables applications in uncertainty quantification, digital twin development, and clinical decision support where traditional 3D simulations are computationally prohibitive. The modular design allows integration with existing 0D solvers, facilitating adoption in established cardiovascular modeling workflows.

Limitations include restriction to rigid-wall assumptions and reliance on synthetic training data that may not capture all physiological variations. Future work should extend the framework to fluid-structure interaction problems, adopt a more complex description of bifurcation geometries to capture the more complex, patient-specific anatomies, and explore extensions of our bifurcation handling method to vessels and other elements of the vasculature.

This hybrid approach demonstrates that strategic integration of data-driven techniques with physics-based numerical methods can significantly advance computational biomedical engineering, providing a pathway toward accurate, efficient, and clinically applicable cardiovascular flow modeling tools.

\section{Acknowledgments}
We would like to thank the National Science Foundation Graduate Research Fellowship Program and Stanford Graduate Fellowship for providing the funding that supported this work.
\section{CRediT Statement}
\textbf{Natalia L. Rubio}: conceptualization, data curation, formal analysis, investigation, methodology, software, validation, visualization, writing – original draft. \textbf{ Eric F. Darve}: conceptualization, formal analysis, supervision, writing – review and editing.  \textbf{Alison L. Marsden}: conceptualization, project administration, resources, supervision, writing – review and editing.

\section{Appendices}

\subsection{3D and 0D Flow Splits}
\begin{figure}[htbp]
    \centering
    \includegraphics[width=0.6\linewidth]{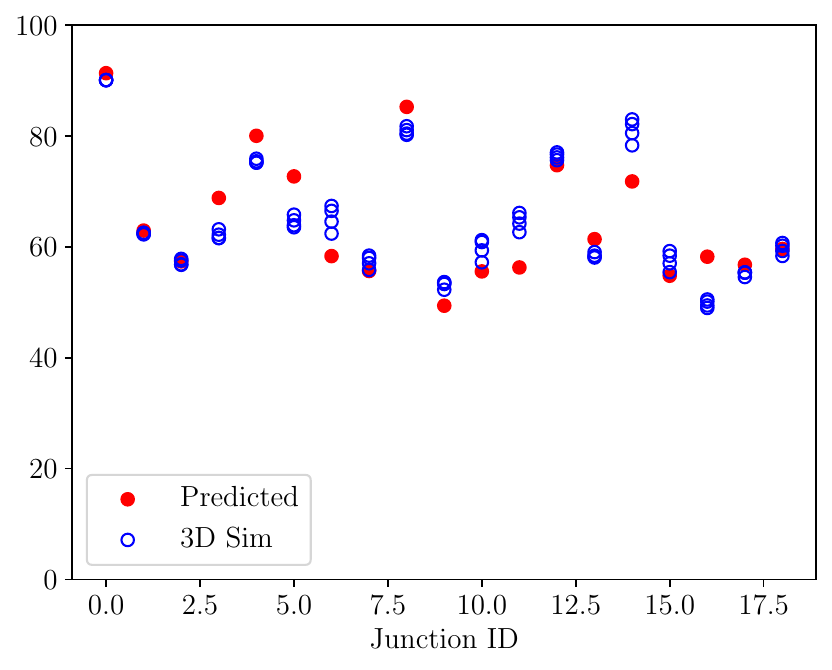}
    \caption{Comparison of flow splits at bifurcations of a 20 outlet tree using \Cref{eq:flow_split} against the flow observed in 3D simulation.}
    \label{fig:flow_splits}
\end{figure}

To test the accuracy of our method of estimating flow splits in a tree before simulation (given by \Cref{eq:flow_split}), we compare the flow splits our method predicts against those observed in 3D simulation in \Cref{fig:flow_splits}, finding good agreement.  This study also confirmed the assumption that the flow split remains relatively constant regardless of the total inflow.
\label{app:flow_splits}
\subsection{Bifurcation Definitions}
\begin{figure}[htbp]
    \centering
    \includegraphics[width=0.99\linewidth]{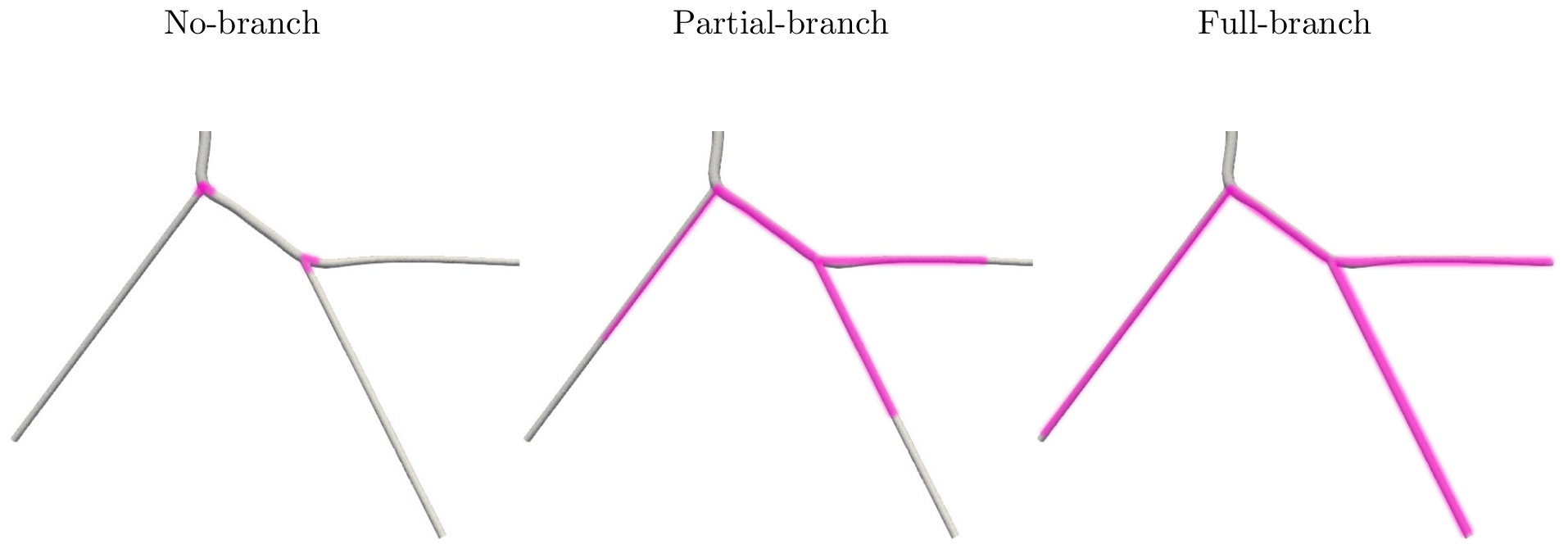}
    \caption{Illustration of three possible bifurcation definitions considered in this work.}
    \label{fig:bif_defs}
\end{figure}

We consider three bifurcation definitions, illustrated in \Cref{fig:bif_defs}, which include the outlet branches to varying degrees.  This work uses the partial-branch definition.  
\label{app:bif_defs}

\subsection{3D Bifurcation Flow and Pressure Fields}
\label{app:3d_fields}
\begin{figure}[htbp]
    \centering
    \includegraphics[width=0.99\linewidth]{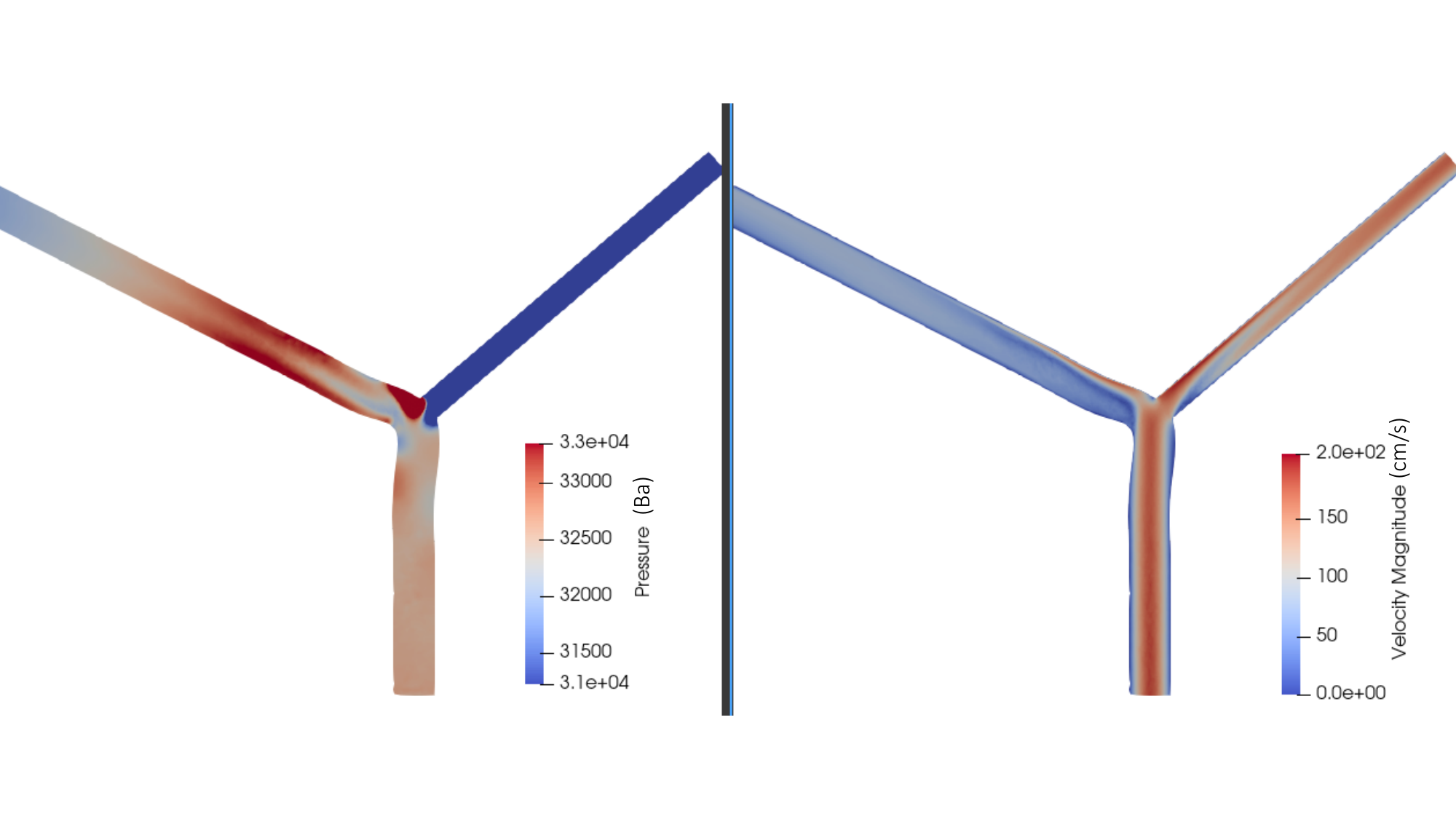}
    \caption{Velocity and pressure fields from a 3D finite-element simulation of a bifurcation under steady flow.}
    \label{fig:3D_bifurcation_fields}
    
\end{figure}

\begin{figure}
    \centering
    \includegraphics[width=0.9\linewidth]{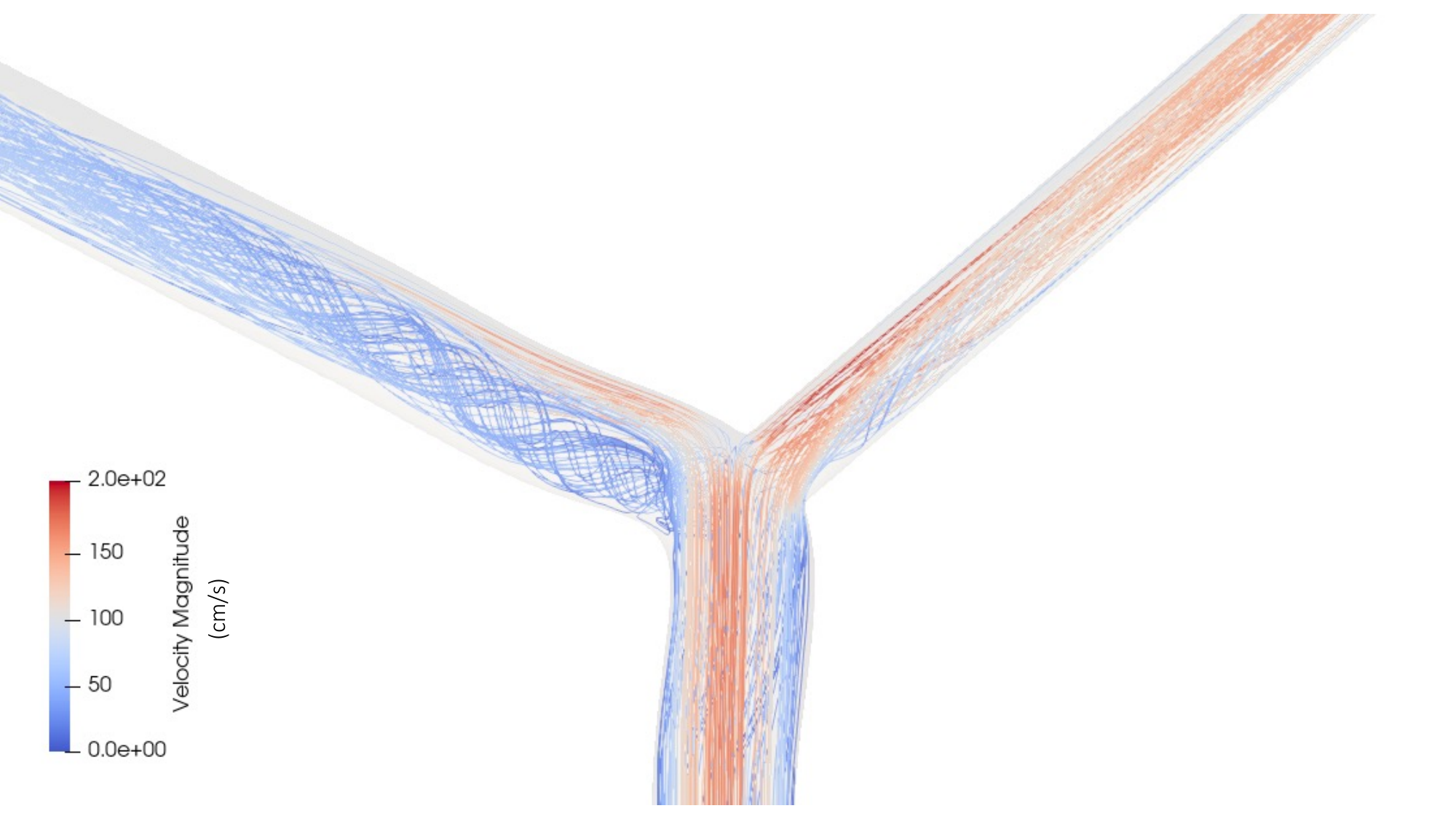}
    \caption{Streamlines in an isolated bifurcation under steady flow.}
    \label{fig:3D_streamlines}
\end{figure}

3D simulations of bifurcations reveal complex flow and pressure effects that are not accounted for in standard ROMs, motivating the development of bifurcation models.  In \Cref{fig:3D_bifurcation_fields}, it is clear that the 0D model assumption of fully-developed Poiseuille flow does not hold - the velocity profile is not parabolic and the pressure does not decrease linearly with distance along the outlet vessel.  \Cref{fig:3D_bifurcation_fields} also motivates the use of the partial-branch bifurcation definition, as it is clear that the bifurcation effects are present well into the outlet branches.
\Cref{fig:3D_streamlines} shows complex flow patterns associated with bifurcations that are not accounted for in reduced-dimensionality ROM formulations.

\subsection{Geometric Feature Ranges}
\label{app:feature_ranges}
We report the number of hidden layers and hidden layer width used in the neural networks that predict each of the RRI and RI coefficients.  Each hidden layer has the same width.
\begin{table}[htpb]
\begin{center}
\begin{tabular}{l c c c}
Parameter &  Minimum & Maximum\\
Daughter 1 area ratio, $\alpha_1$ & 0.40 & 1.2  \\
Daughter 2 area ratio, $\alpha_2$ & 0.37 & 1.2  \\
Daughter 1 length ratio, $\lambda_1$  & 15 & 41 \\
Daughter 2 length ratio, $\lambda_2$ & 16 & 42  \\
Daughter 1 angle (rad), $\theta_1$, & 0.05 & 1.41 \\
Daughter 2 angle (rad), $\theta_2$& 0.33 & 1.51  \\
Daughter 1 flow split, $\phi_1$, & 0.15 & 0.89 \\
Daughter 2 flow split, $\phi_2$& 0.10 & 0.85  \\

\end{tabular}
\caption{Ranges of non-dimensional parameters describing bifurcation geometry.}
\label{tab:feature_ranges}
\end{center}
\end{table}

\subsection{Neural Network Parameters}
\label{app:nn_params}
We report the number of hidden layers and hidden layer width used in the neural networks that predict each of the RRI and RI coefficients.  Each hidden layer has the same width.  We train a separate model for each coefficient, and the two outlet types (1) constructed as a continuation of the parent vessel, and (2) constructed as a separate vessel and then unioned with the parent vessel.
\begin{table}[htpb]
\begin{center}
\begin{tabular}{l c c c}
Coefficient &  \# Hidden Layers & Layer Width\\
Outlet 1 RRI Linear Resistance & 2 & 15  \\
Outlet 2 RRI Linear Resistance & 1 & 40  \\
Outlet 1 RRI Quadratic Resistance & 2 & 30  \\
Outlet 2 RRI Quadratic Resistance & 1 & 23  \\
Outlet 1 RRI Inductance & 1 & 12 \\
Outlet 2 RRI Inductance & 1 & 12 \\
Outlet 1 RI Linear Resistance & 1 & 10  \\
Outlet 2 RI Linear Resistance & 1 & 20  \\
Outlet 1 RI Inductance & 1 & 20  \\
Outlet 2 RI Inductance & 1 & 40  \\

\end{tabular}
\caption{Number of hidden layers and layer width for each neural network model used to predict RRI or RI coefficients.}
\label{tab:nn_params}
\end{center}
\end{table}

\newpage
\bibliographystyle{abbrvnat}
\bibliography{main}
\end{document}